\documentclass[runningheads]{llncs}
\usepackage{graphicx}
\usepackage{amssymb}
\usepackage[table]{xcolor}
\usepackage{multirow}
\usepackage{amsmath} 

\definecolor{lightred}{rgb}{1,0.9,0.9}
\definecolor{lightblue}{rgb}{0.9,0.9,1.0}
\definecolor{red}{rgb}{1,0,0}

\begin{document}
\title{Arges: Spatio-Temporal Transformer for Ulcerative Colitis Severity Assessment in Endoscopy Videos}
%
%
%
%
%
\titlerunning{Arges}
\author{Krishna Chaitanya \inst{1} \and
Pablo F. Damasceno \inst{1} \and
Shreyas Fadnavis \inst{1} \and
Pooya Mobadersany \inst{1} \and
Chaitanya Parmar \inst{1} \and
Emily Scherer \inst{1} \and
Natalia Zemlianskaia \inst{1} \and
Lindsey Surace \inst{1} \and
Louis R. Ghanem \inst{1} \and
Oana Gabriela Cula \inst{1} \and
Tommaso Mansi \inst{1} \and
Kristopher Standish \inst{1}
}

\authorrunning{Chaitanya, K. et al.}

\institute{$^{1}${Janssen R\&D, LLC, a Johnson \& Johnson Company}\\
\email{kchaita6@its.jnj.com}}
\maketitle              
\begin{abstract}
Accurate assessment of disease severity from endoscopy videos in ulcerative colitis (UC) is crucial for evaluating drug efficacy in clinical trials. Severity is often measured by the Mayo Endoscopic Subscore (MES) and Ulcerative Colitis Endoscopic Index of Severity (UCEIS) score. However, expert MES/UCEIS annotation is time-consuming and susceptible to inter-rater variability, factors addressable by automation. 
Automation attempts with frame-level labels face challenges in fully-supervised solutions due to the prevalence of video-level labels in clinical trials. CNN-based weakly-supervised models (WSL) with end-to-end (e2e) training lack generalization to new disease scores and ignore spatio-temporal information crucial for accurate scoring. 
To address these limitations, we propose ``Arges", a deep learning framework  that utilizes a transformer with positional encoding to incorporate spatio-temporal information from frame features to estimate disease severity scores in endoscopy video. Extracted features are derived from a foundation model (ArgesFM), pre-trained on a large diverse dataset from multiple clinical trials (61M frames, 3927 videos). 
We evaluate four UC disease severity scores, including MES and three UCEIS component scores. Test set evaluation indicates significant improvements, with F1 scores increasing by 4.1\% for MES and 18.8\%, 6.6\%, 3.8\% for the three UCEIS component scores compared to state-of-the-art methods. Prospective validation on previously unseen clinical trial data further demonstrates the model's successful generalization.

\end{abstract}
%
%
\section{Introduction}
Ulcerative colitis (UC), a chronic inflammatory bowel disease (IBD), impacts approximately 5 million individuals worldwide, causing intestinal inflammation and ulceration. In UC clinical trials, colon disease severity is often assessed through endoscopy videos and measured using standard scoring systems such as Mayo Endoscopic Subscore (MES)\cite{schroeder1987coated} and  Ulcerative Colitis Endoscopic Index of Severity (UCEIS)\cite{travis2012developing}. 
Expert human assessment of videos 
is time-consuming and prone to inter-rater variability, emphasizing the need for automated solutions. Automating endoscopic disease scoring, however, presents unique challenges: 1) disease scoring is not common practice in clinical settings, so annotated datasets are scarce; 2) in clinical trials, where data is routinely annotated, labeling is performed at the video level, challenging the use of readily-available, frame-based fully-supervised networks; 3) because videos are typically long, reaching over 30 minutes in length, full video frame-wise annotation and use of full videos as input for 3D CNNs and LSTM is challenging. In addition, recent findings\cite{rubin2023development} suggest that temporal awareness is an important component for manual and algorithmic disease scoring, signaling the need to move beyond static, frame-based models.

Significant progress has been made towards accurate disease assessment in endoscopy videos with fully-supervised networks, trained on frame-level annotations\cite{stidham2019performance,vasilakakis2020weakly,stidham2024using}. 
These works are particularly well-suited for detection of single, easier to annotate features such as polyps but given the high cost and complexity of frame-level annotation in UC, no large and public datasets exist to date, hampering the creation of accurate and generalizable models for MES/UCEIS scoring. In UC clinical trials, where annotated data is abundant, labeling is typically reported on the video level and techniques such as weakly-supervised Learning (WSL) can be powerful. Initially used for natural images\cite{paul2018w,wang2017untrimmednets} adaptation of WSL to medical imaging modalities such as endoscopy\cite{schwab2022automatic,polat2022class,tian2022contrastive,byrne2023application,iacucci2023virtual} has been prominent. Recently, a WSL solution using CNNs and multiple instance learning (MIL)\cite{ilse2018attention} was introduced\cite{schwab2022automatic}, demonstrating state-of-the-art (SOTA) performance in assessing UC disease severity in endoscopy videos. A subsequent extension\cite{polat2022class} replaced this work's ordinal loss with a distance-weighted loss to penalize larger misclassifications and improve model accuracy.

While accurate, end-to-end (e2e) training of these CNN-based WSL models are computationally costly. Recent works\cite{wang2023foundation,hirsch2023self,xu2022patch} attempted to overcome this by training a foundation encoder, emphasizing generalizability through self-supervised training (SSL)\cite{caron2021emerging,oquab2023dinov2}. These versatile models demonstrate robust performance in various downstream tasks, including surgical phase recognition, polyp and lesion detection, and patch-level disease characterization in IBD and are therefore prime candidates for accurate and inexpensive UC disease scoring. However, long-range spatio-temporal modeling remains lacking in both WSL and SSL models, a feature that we hypothesize to be crucial for accurate disease severity scoring from endoscopy videos in UC.

To address these drawbacks, we introduce ``Arges", a robust and generalizable deep learning framework for UC disease characterization from endoscopy videos. It comprises of a foundational SSL encoder (ArgesFM) followed by downstream classifier (ArgesMES/ArgesUCEIS) to estimate disease severity score per video. 
Mimicking clinical experts workflow, the downstream classifier is specifically designed to model spatio-temporal information by the use of transformers with positional embeddings. Impractical computation costs of long-range temporal modeling are avoided by using low-dimensional features as inputs, obtained from a foundational model (ArgesFM) trained using SSL on a large diverse data.

Our key contributions are summarized as follows: 
\begin{itemize}
    \item \textbf{Spatio-Temporal Modeling:} Incorporating spatio-temporal information with a transformer-based classifier via positional encoding, mirroring clinical experts intuition, leads to performance gains in F1 score over SOTA methods (4.1\% on MES, and 18.8\%, 6.6\%, 3.8\% on three UCEIS component scores).
    \item \textbf{Robust foundation encoder:} A reliable foundation encoder (ArgesFM) is crucial for efficiently training downstream models. ArgesFM, trained on over 61M frames from diverse clinical trial data, produces generalizable features essential for developing diverse disease severity models.
    \item \textbf{Prospective validation:} Demonstrate successful generalizability to a multi-site, international prospective clinical trial with unseen data (14M frames).
    \item \textbf{Large, Diverse Data Curation:} Our dataset spans four multi-site clinical trials across continents, covering 2 IBD subtypes and various severities. It's the largest IBD dataset for SSL pre-training, 14x larger than previous model.
\end{itemize}

\section{Data and Methods}
\begin{figure}[!t]
    \centering    \includegraphics[width=1.1\linewidth]{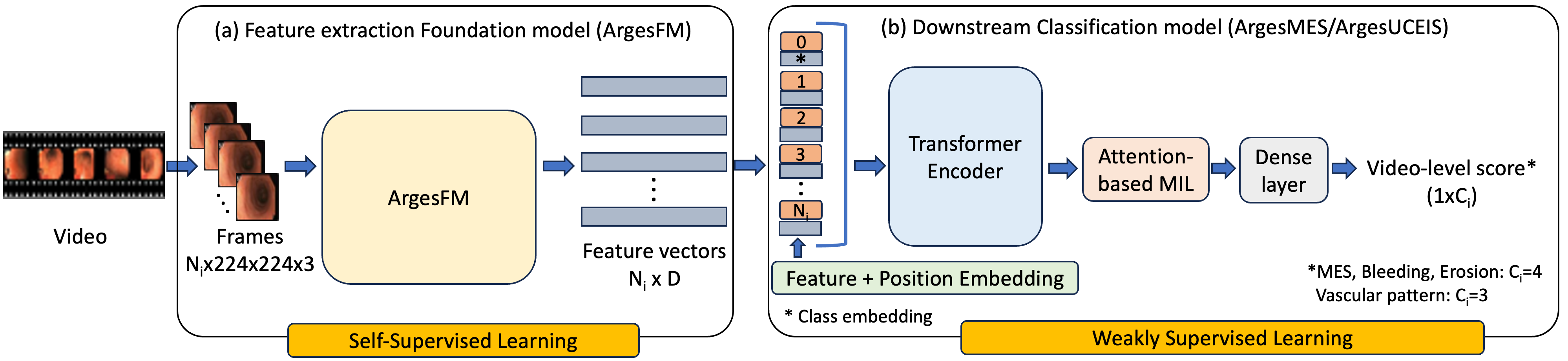}
    \caption{Arges framework has a foundation model (ArgesFM) followed by a downstream classifier (ArgesMES/ArgesUCEIS) to estimate disease severity score per video.}
    \label{fig:methods}
\end{figure}

Our framework as in Fig.\ref{fig:methods} is comprised of two components: Foundation model for feature extraction from frames (Sec 2.2), pre-trained with curated data (Sec 2.1), and a downstream classification model (Sec 2.3) to estimate severity score.

\textbf{2.1 IBD disease data curation:}
Inflammatory Bowel Diseases (IBD), including Ulcerative Colitis (UC) and Crohn's Disease (CD) are chronic gastrointestinal disorders characterized by inflammation, ulcers, and rectal bleeding.
In our study, we use endoscopy videos from four clinical trials (two UC\cite{sands2018peficitinib,sands2019ustekinumab}, two CD\cite{sands2022ustekinumab,allez2023phase}) focused on drug safety and efficacy for moderate to severe UC or CD as shown in Table \ref{table1:datasets_new}. 
This diverse dataset, spanning 5 continents and 30 countries, comprised 2411 patients, 4911 videos, and over 71M frames. The UC trials data included video-wise labels for MES and UCEIS subscores.
For foundational encoder pre-training, we used 61M frames, a substantial 14x increase compared to previous largest SSL model for endoscopy videos\cite{wang2023foundation}. 
For downstream tasks, unlike\cite{wang2023foundation} which uses 8-frame clips for simpler tasks like polyp detection, our approach tackles a complex UC disease severity scoring task, requiring long-range modeling for accurate evaluation. Our ablation study explores the impact of clip- versus full-video pre-training.
Our downstream models were rigorously evaluated on an unseen prospective dataset\cite{peyrin2023guselkumab} from a multi-center, international UC clinical trial (14M frames).

\begin{table}[!t]
\begin{tabular}{|>{\centering\arraybackslash}p{1.9cm}|>{\centering\arraybackslash}p{0.8cm}|>{\centering\arraybackslash}p{1.2cm}|>{\centering\arraybackslash}p{1cm}|>{\centering\arraybackslash}p{1.3cm}|>{\centering\arraybackslash}p{2.2cm}|>{\centering\arraybackslash}p{2.2cm}|>{\centering\arraybackslash}p{2.2cm}|}
\hline
Dataset & IBD (N) & Patients (N) & Videos (N) & Frames (million) & Foundation model training & Downstream model training & Downstream model testing \\ \hline
SEAVUE\cite{sands2022ustekinumab} & CD & 371 & 631 & 8.2 & \textbf{Yes} (80\%) & No & No \\ \hline
TRIDENT\cite{allez2023phase} & CD & 382 & 704 & 13.1 & \textbf{Yes} (80\%) & No & No\\ \hline
\rowcolor{lightblue}UNIFI\cite{sands2018peficitinib} & UC & 1,105 & 3,128 & 42 & \textbf{Yes} (80\%) & \textbf{Yes} (80\%) & Yes (20\%)\\ \hline
\rowcolor{lightblue}JAKUC\cite{sands2019ustekinumab} & UC & 286 & 448 & 8.1 & \textbf{Yes} (80\%) & \textbf{Yes} (80\%) & Yes (20\%) \\ \hline 
\rowcolor{lightred}QUASAR\cite{peyrin2023guselkumab} (Prospective)& UC & 313 & 615 & 14 & No & No & \textbf{Yes} (100\%) \\ \hline
\end{tabular}
\caption{Overview of UC and CD datasets. The first 4 datasets are used for ArgesFM pre-training. UC datasets in blue are for downstream models (ArgesMES/ArgesUCEIS) training/testing. UC dataset in red is used for prospective validation.}
\label{table1:datasets_new}
\end{table}


\textbf{2.2 Foundation model (ArgesFM):} To leverage unlabeled data from diverse IBD trials, we developed a foundation model (FM) based on SSL\cite{oquab2023dinov2}, outlined in Fig.\ref{fig:methods}. ArgesFM was trained on both UC and CD data from four clinical trials to enhance generalizability to unseen data and adaptability to various downstream tasks without requiring end-to-end training. 
ArgesFM utilizes a vision transformer (ViT-Base\cite{dosovitskiy2020image}, 86M parameters) to encode video frames, capturing intricate spatial relationships within a frame through self-attention, trained using  DINOv2\cite{oquab2023dinov2}. DINOv2 employs a student-teacher model paradigm, where both models share the same architecture in a knowledge distillation setup. The student model is exposed to multi-crop variants of input, including global views (224x224x3) and local views (96x96x3), while the teacher model is exposed to only global views. The student model is optimized to match teacher model representations. ArgesFM undergoes training using a combined loss function, $L_{\text{DINOv2}} = L_{\text{DINO}} + L_{\text{iBOT}} + L_{\text{koleo}}$, which includes DINO loss, iBOT loss, and KoLeo regularization, as proposed in [7]. Teacher model outputs undergo centering via batch mean, with weights ($\theta_t$) updated using an exponential moving average (EMA) of the student model weights ($\theta_s$).

\textbf{2.3 Downstream classifier (ArgesMES or ArgesUCEIS):} When evaluating disease severity in endoscopy videos, clinicians prefer whole videos or clips\cite{rubin2023development} over individual frames\cite{stidham2019performance,schwab2022automatic}, as context from multiple frames helps distinguish true disease features from artifacts like camera blur or forceps-induced bleeding. To test if such dynamic content improves disease quantification accuracy, we employed a compact temporal model (Transformer\cite{vaswani2017attention}, 17M parameters) to estimate disease severity from videos pre-encoded by ArgesFM. In this design, an input matrix of $N \times D$ features is obtained by passing a video with $N$ frames through the trained student network, $D$ is feature vector dimension ($D=768$). The extracted features undergo positional encoding before being input into a Transformer, followed by an attention-based MIL aggregator and a dense layer to estimate a severity score for each video. This novel approach, distinct from prior SOTA models, effectively captures temporal dynamics by incorporating positional encoding and has the potential to improve performance. To capture temporal information, we use positional encoding (PE) for the input matrix $N \times D$: $\text{PE}(\text{pos}, 2i) = \sin\left(\frac{\text{pos}}{10000^{2i/D}}\right)$ and $\text{PE}(\text{pos}, 2i+1) = \cos\left(\frac{\text{pos}}{10000^{2i/D}}\right)$ as in\cite{vaswani2017attention}. Here, $i$ represents the dimension index.
Our model uses a multi-head attention mechanism for analyzing temporal sequences, assigning weights to frames via: $\text{Attention}(Q, K, V) = \text{softmax}\left(\frac{QK^T}{\sqrt{d_k}}\right)V$, where $Q$, $K$, $V$, and $d_{k}$ denote queries, keys, values, and dimension of key vectors $(Q, K, V)$. 
Multi-head attention captures inter-frame correlations, potentially enhancing classification accuracy: $\text{MultiHead}(Q, K, V) = \text{Concat}(\text{head}_1, \text{head}_2, \ldots, \text{head}_h)W^O$, where each head is computed as: $\text{head}_i = \text{Attention}(QW_i^Q, KW_i^K, VW_i^V)$, with $W_i^Q$, $W_i^K$, $W_i^V$, and $W^O$ being learned projection matrices. The attention-based MIL aggregator is chosen for its compatibility with video-level labels in a weakly supervised learning (WSL) setting, providing ``high-attention" regions that facilitate clinical interpretation and model quality control. Leveraging shared features from our foundation model, we efficiently train different disease severity models using our proposed compact downstream model architecture. Unlike WSL methods requiring full end-to-end training, our approach enables fast training. We trained four independent downstream models: ArgesMES for MES scores (0-3) and three models for UCEIS component scores (ArgesUCEIS): bleeding (0-3), erosion (0-3), and vascular pattern (0-2).

\section{Data splits, benchmarking, and implementation details}

\textbf{Dataset split:} 
We partitioned data from two UC and two CD clinical trials into training (80\%) and held-out test (20\%) sets.
ArgesFM was trained on this 80\% training data, which consisted of over 61M frames. The downstream models (ArgesMES/ArgesUCEIS) require MES and UCEIS scores annotations, available only for UC trials. Hence, we trained the downstream model using the same 80\% training data from two UC trials (UNIFI, JAKUC) through 4-fold cross-validation. 
Performance evaluation is reported on the held-out test set (20\%) for UNIFI and JAKUC. Prospective validation was conducted on an unseen third UC trial (QUASAR) (100\% data) using each locked model from individual training folds (Table 1).

\textbf{Data Pre-processing:} Videos were converted to frames at 30 fps, resized to 224x224 (native video resolution varied between 640x510 to 1280x960) and 3 RGB channels normalized using ImageNet values.

\textbf{Arges models architecture:} For ArgesFM, we employ ViT-Base\cite{dosovitskiy2020image} as the encoder for SSL training, utilizing DINOv2 and it acts as a feature extractor (first block in Fig. 1). Downstream models (second block in Fig. 1): ArgesMES or ArgesUCEIS incorporates a compact transformer with two encoder layers, each equipped with four multi-attention heads and dropout of 0.25. This is followed by an attention-based MIL and a dense layer classifier with dropout of 0.5 to estimate a disease severity score per video. 

\textbf{Training Details:} ArgesFM (ViT-Base) was trained for 300k iterations on 4 A10G GPUs using DINOv2 and parameters\cite{oquab2023dinov2}, using 1 global crop (224x224x3) and 8 local crops (96x96x3) with a batch size of 256. 
For downstream tasks, we independently trained four downstream classifier models to estimate MES (ArgesMES) and three UCEIS (ArgesUCEIS) component scores (bleeding, erosion, and vascular pattern) for 15 epochs with a learning rate of $10^{-4}$ and weight decay of $10^{-5}$ on 1 A10G GPU.
We used multiclass cross-entropy loss and weighted sampling to address class imbalance in the data. 

\textbf{Comparison with state-of-the-art:} For WSL baseline comparison, we used ResNet34 encoders as in\cite{schwab2022automatic,polat2022class}. Max pooling and Attention-based MIL\cite{schwab2022automatic} aggregator models were trained for 150 epochs. The CDW-CE\cite{polat2022class} model with additional novel loss on top of\cite{schwab2022automatic} with attention-based MIL was trained for 150 epochs. For SSL models, we used publicly available EndoFM\cite{wang2023foundation} pre-trained SSL model weights to extract features, pre-trained with colonoscopy, laparoscope, and gastroscope data using DINOv1\cite{caron2021emerging}, and trained only the downstream classifier for 25 epochs as recommended in\cite{wang2023foundation} to estimate disease severity score per video. 

\textbf{Evaluation Metrics:} Performance assessment was done using a F1 weighted scores evaluation based upon ground-truths from clinical trial annotators (central reader). Additionally, weighted Cohen Kappa was used to measure the agreement between the proposed downstream classifier and human readers and compared to inter-reader agreement between two human readers.

\section{Experiments, Results and Discussion}

\begin{table}[!t]
  \centering
  \begin{tabular}{|p{5.5cm}|>{\columncolor{lightblue}\centering\arraybackslash}p{2.3cm}|>{\columncolor{lightblue}\centering\arraybackslash}p{2.3cm}|>{\columncolor{lightred}\centering\arraybackslash}p{2.8cm}|}
    \hline
    \textbf{Method (aggregation)} & \textbf{Test Set (UNIFI)} & \textbf{Test Set (JAKUC)} & \cellcolor{lightred}\textbf{Prospective data (QUASAR)} \\ \hline
    WSL\cite{schwab2022automatic} CNNs (Max pooling) & 0.550 ± 0.01 & 0.458 ± 0.05 & 0.632 ± 0.01 \\ \hline
    WSL\cite{schwab2022automatic} CNNs (Attention) & 0.603 ± 0.01 & 0.585 ± 0.03 & 0.633 ± 0.03 \\ \hline
    CDW-CE\cite{polat2022class} CNNs (Attention) & 0.592 ± 0.01 & 0.599 ± 0.03 & 0.605 ± 0.01  \\ \hline
    EndoFM\cite{wang2023foundation} $\rightarrow$ LC (Avg. pooling) & 0.419 ± 0.01 & 0.285 ± 0.02 & 0.501 ± 0.01 \\ \hline
    ArgesFM  $\rightarrow$ LC (Avg. pooling) & 0.548 ± 0.01 & 0.442 ± 0.01 & 0.609 ± 0.02 \\ \hline
    \textbf{ArgesFM $\rightarrow$ T (Attention)} & \textbf{0.644 ± 0.01}$^{\ast}$ & \textbf{0.620 ± 0.01}$^{\ast}$ & \textbf{0.698 ± 0.01}$^{\ast}$\\ \hline
  \end{tabular}
  \caption{Benchmarking ArgesMES against SOTA models shows significant performance boosts on test set and prospective data for MES classification. LC: Linear classifier, T: Transformer. Wilcoxon test (p$<$0.05) shows statistical significance ($\ast$).}
  \label{table2:mes_model}
\end{table}
\textbf{4.1. Arges outperforms state-of-the-art (SOTA) models on MES classification:} Table 2 compares Arges with SOTA models, including WSL and SSL, on MES classification across two UC held-out test sets (UNIFI, JAKUC) and prospective trial data (QUASAR). WSL\cite{schwab2022automatic} exhibits similar trends as reported previously, where Attention MIL and a modified loss outperform a basic max pooling model. 
EndoFM\cite{wang2023foundation}, a comparable foundation model, although effective in simpler tasks like polyp detection with short clips of 8 frames, shows poor generalization to unseen UC disease types and yields low F1 scores for MES classification, likely due to the need for a larger range of frames for accurate severity assessment. 
To directly compare with EndoFM, we incorporated average pooling into Arges, which resulted in better performance compared to EndoFM, but did not show any significant F1 improvement when compared to WSL max pooling.
In contrast, utilizing a Transformer network with attention MIL for the downstream task, rather than a simpler aggregation, demonstrates a notable F1 score increase of 4.1\% and 2.1\% in UNIFI and JAKUC test sets over SOTA models. (More details for ablation experiments in section 4.4).

\textbf{4.2. All models significantly generalize to unseen, prospective data:} Four comparison models and our two new models demonstrate non-inferior F1 scores on the unseen QUASAR dataset (refer to Table 2 and 3). This outcome is attributed to the extensive and diverse datasets used for model training, enabling effective generalization to new data.

\textbf{4.3. Arges excels on other downstream scoring tasks without e2e training:} Table 3 presents the results of employing ArgesFM features for other disease severity tasks, specifically scoring the three UCEIS components (bleeding, erosion, vascular pattern). A comparison with the WSL CNNs attention-MIL\cite{schwab2022automatic} model (achieves the highest F1 score in Table 2) demonstrates notable performance gains with our method when ArgesFM is used with Transformer and attention MIL, surpassing CNN-based WSL. Moreover, these models exhibit improved training and inference time on unseen data, detailed in Supplementary.

\begin{table}[!t]
  \centering
  \begin{tabular}{|>{\centering\arraybackslash}p{2.1cm}|p{2.5cm}|>{\columncolor{lightblue}\centering\arraybackslash}p{2.4cm}|>{\columncolor{lightblue}\centering\arraybackslash}p{2.4cm}|>{\columncolor{lightred}\centering\arraybackslash}p{2.9cm}|}
    \hline
    \textbf{Target UCEIS score} & \textbf{Method} & \textbf{Test Set (UNIFI)} & \textbf{Test Set (JAKUC)} & \cellcolor{lightred}\textbf{Prospective data (QUASAR)} \\ \hline
    \multirow{2}{*}{Bleeding} & WSL\cite{schwab2022automatic} CNNs & 0.434 ± 0.04 & 0.334 ± 0.04 & \textbf{0.444 ± 0.03} \\ \cline{2-5}
    & \textbf{Arges} & \textbf{0.624 ± 0.02}$^{\ast}$ & \textbf{0.525 ± 0.03}$^{\ast}$ & 0.384 ± 0.01 \\ \hline
    \multirow{2}{*}{Erosion} & WSL\cite{schwab2022automatic} CNNs & 0.545 ± 0.01 & 0.488 ± 0.04 & 0.437 ± 0.02 \\ \cline{2-5}
     & \textbf{Arges} & \textbf{0.611 ± 0.02}$^{\ast}$ & \textbf{0.637 ± 0.01}$^{\ast}$ & \textbf{0.446 ± 0.02}$^{\ast}$ \\ \hline
     {Vascular} & WSL\cite{schwab2022automatic} CNNs & 0.693 ± 0.02 & 0.602 ± 0.04 & 0.704 ± 0.02 \\ \cline{2-5}
     {pattern} & \textbf{Arges} & \textbf{0.731 ± 0.01}$^{\ast}$ & \textbf{0.635 ± 0.05}$^{\ast}$ & \textbf{0.723 ± 0.01}$^{\ast}$ \\ \hline
  \end{tabular}
  \caption{ArgesUCEIS includes three models for UCEIS scores (bleeding, erosion, vascular pattern). Using ArgesFM with ArgesUCEIS outperforms WSL CNNs with attention MIL\cite{schwab2022automatic}. Wilcoxon test (p $<$ 0.05) shows statistical significance, denoted by $\ast$.}
  \label{table3:uceis_models}
\end{table}

\begin{table}[!t]
  \centering
  \begin{tabular}{|p{4.0cm}|>{\centering\arraybackslash}p{0.8cm}|>{\centering\arraybackslash}p{0.8cm}|>{\columncolor{lightblue}\centering\arraybackslash}p{2.2cm}|>{\columncolor{lightblue}\centering\arraybackslash}p{2.2cm}|>{\columncolor{lightred}\centering\arraybackslash}p{2.75cm}|}
    \hline
    \textbf{Method (MES scoring)} & \textbf{T} & \textbf{Attn} & \textbf{Test Set (UNIFI)} & \textbf{Test Set (JAKUC)} & \cellcolor{lightred}\textbf{Prospective data (QUASAR)} \\ \hline
    ArgesFM $\rightarrow$ Avg. pooling & $\times$ & $\times$ & 0.548 ± 0.01 & 0.442 ± 0.01 & 0.609 ± 0.02 \\ \hline
    ArgesFM $\rightarrow$ T & $\checkmark$ & $\times$ & 0.626 ± 0.01 & 0.593 ± 0.04 & 0.616 ± 0.02 \\ \hline
    ArgesFM $\rightarrow$ Attn-MIL & $\times$ & $\checkmark$ & 0.622 ± 0.01 & 0.612 ± 0.01 & 0.662 ± 0.02 \\ \hline
    ArgesFM $\rightarrow$ T + Attn-MIL & $\checkmark$ & $\checkmark$ & \textbf{0.644 ± 0.01}$^{\ast}$ & \textbf{0.620 ± 0.01}$^{\ast}$ & \textbf{0.698 ± 0.01}$^{\ast}$ \\
    \hline
  \end{tabular}
  \caption{Ablation experiments evaluated network components' impact in the downstream classifier, activating ($\checkmark$) or deactivating ($\times$) elements like transformer (T) and attention-based MIL (Attn-MIL). Wilcoxon test (p $<$ 0.05), statistical significance ($\ast$).}
  \label{table4:ablation_exps}
\end{table}

\textbf{4.4. MIL aggregator and dynamics of Transformer improve disease scoring performance:} 

Table \ref{table2:mes_model} shows enhanced disease scoring accuracy with the Transformer architecture compared to simple average pooling. In Table \ref{table4:ablation_exps}, we detail our ablation study on the contributions of Transformer and MIL aggregator components. Initially, integrating a Transformer alone improved performance, emphasizing the importance of long-range spatio-temporal modeling that aggregates insights from consecutive frames. Subsequent addition of an attention MIL aggregator alone boosted performance, leveraging its effectiveness in weakly-supervised scenarios. Optimal results were achieved with both components, highlighting their collaborative effectiveness.

\textbf{4.5. ArgesMES scores are interchangeable with human readers:} Previous studies\cite{daperno2014inter,principi2020inter} has highlighted the inherent subjectivity in scoring UC severity by experts and quantify it using the weighted Cohen's kappa score (k).
We had two expert readers MES evaluations with the prospective trial data.
In the prospective data evaluation, the agreement between ArgesMES output and human reader assessment (k=0.66, CI=0.60–0.72) closely aligns with the two human expert raters agreement (k=0.71, CI=0.66–0.76). This falls within the range (k=0.61 to 0.8) considered as substantial agreement, as in prior studies\cite{schwab2022automatic,daperno2014inter,principi2020inter}.

\begin{figure}[!t]
    \centering
    \includegraphics[width=1.0\linewidth]{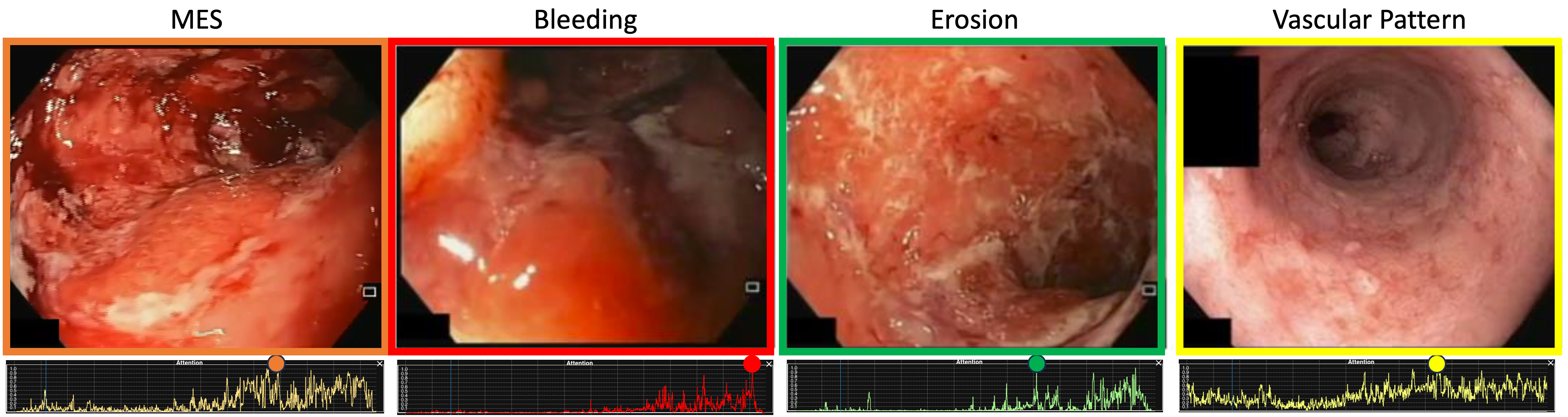}
    \caption{Representative frames from ArgesMES $\&$ ArgesUCEIS models display high-attention regions along with respective attention maps for all video frames at bottom.}
    \label{fig:interpretability}
\end{figure}

\textbf{4.6. Interpretability:} By using Attention-based MIL, attention scores are obtained for each frame in the video. Fig.\ref{fig:interpretability} shows the high attention regions where the model focuses, influencing the determination of severity score. In a qualitative assessment, two independent experienced gastroenterologists examined these attention regions in 15 videos. 
Their consensus was that the model prioritizes informative areas, aligning with clinical intuition. 
This design improves interpretability and may assist in making informed decisions during clinical trials.

\section{Conclusion}
The ``Arges" framework presents key advancements in endoscopy video analysis. The incorporation of spatio-temporal modeling, utilizing a transformer with positional encoding, yields notable performance gains across four distinct severity scoring models: ArgesMES for automating MES and ArgesUCEIS for three UCEIS component scores. The robust foundational encoder, ArgesFM, trained on a large and diverse clinical trial dataset, yields generalizable features, facilitating efficient downstream model training for various disease severity tasks. Successful prospective validation on an unseen clinical trial further affirms the framework's effectiveness. Furthermore, the implementation of attention-based MIL enhances interpretability, highlighting ``high-attention" regions for quality control checks by clinical experts during model development and drug research.\\

\textbf{Disclosure of Interests.} All authors were employees of Janssen R\&D, LLC, and may own company stock/stock options.

%
%
%
\bibliographystyle{splncs04}
\bibliography{Paper-012}

\clearpage 

\section{Supplementary}

\begin{table}
  \centering
  \begin{tabular}{|p{3.6cm}|p{4.9cm}|p{3.5cm}|}
    \hline
    \textbf{} & \textbf{Method} & \textbf{Time} \\ \hline
    \textbf{Training} on & CNNs-based WSL [17] (Training) & $\sim$ 24 hours\\ \cline{2-3}
    UC trial data (80\%) & \textbf{ArgesFM} (Pre-training) &  $\sim$ 8 days  \\ \cline{2-3}
    (UNIFI, JAKUC) & \textbf{ArgesMES} (Training) & $\sim$ 1 hour\\ \hline
    \textbf{Inference} on  & {CNNs-based WSL [17] (Inference)}&  $\sim$ 21 hours \\ \cline{2-3}
    Prospective, unseen trial & \textbf{ArgesFM} (Feature extraction) &  $\sim$ 4 hours \\ \cline{2-3}
    data (QUASAR) & \textbf{ArgesMES} (Inference) & $\sim$ 10 minutes\\ \hline
  \end{tabular}
  \caption{We present the runtime analysis results in this table. The first three rows display the time required to train the following models: CNN-based weakly supervised learning (WSL) methods, Foundation model (ArgesFM), and downstream classifier model (ArgesMES). Similarly, the last three rows demonstrate the inference time required for all models on a prospective, unseen trial dataset (QUASAR)}
\end{table}


\clearpage 





\end{document}